\documentclass[aps,prl,twocolumn, showpacs]{revtex4-1}
\pdfoutput=1
\usepackage[latin9]{inputenc}
\usepackage{amstext}
\usepackage{graphicx}
\usepackage{amssymb}
\usepackage{esint}
\begin{document}

\title{The effect of Landau-Zener dynamics on phonon lasing}

\author{Huaizhi Wu$^{1,2}$, Georg Heinrich$^1$ and Florian Marquardt$^{1,3}$}

\affiliation{$^1 $ Institute for Theoretical Physics, Universit\"at Erlangen-N\"urnberg, Staudtstr.
7, 91058 Erlangen, Germany}

\affiliation{$^2 $ Department of Physics, Fuzhou University, Fuzhou 350002, People's Republic of China}

\affiliation{$^3 $ Max Planck Institute for the Science of Light, G\"unter-Scharowsky-Stra�e
1/Bau 24, 91058 Erlangen, Germany}


\begin{abstract}
Optomechanical systems couple light to the motion of nanomechanical
objects. Intriguing new effects are observed in recent experiments
that involve the dynamics of more than one optical mode. There, mechanical
motion can stimulate strongly driven multi-mode photon dynamics that
acts back on the mechanics via radiation forces. We show that even
for two optical modes Landau-Zener-Stueckelberg oscillations of the
light field drastically change the nonlinear attractor diagram of
the resulting phonon lasing oscillations. Our findings illustrate
the generic effects of Landau-Zener physics on back-action induced
self-oscillations.
\end{abstract}



\maketitle
The exploration of nanomechanical objects and their interaction with
light constitutes the rapidly evolving field of optomechanics (see
\cite{2012_Aspelmeyer_QuOptomechanics,2013_AspelmeyerKippenbergMarquardt_OptomechanicsReview}
for recent reviews). The key element of any optomechanical system
is a laser-driven optical mode whose resonance frequency shifts in
response to the displacement of a mechanical object. The photon dynamics
conversely acts back on the mechanics in terms of a radiation pressure
force. These dynamical back-action effects, mediated by the light
field, can cool or amplify mechanical motion, and even drive the system
into a regime of self-induced mechanical oscillations \cite{2004_KarraiConstanze_IEEE,2005_Rokhsari_OptExp,Carmon2005Temporal-Behavi,Kippenberg2005Analysis-of-Rad,Marquardt2006Dynamical-Multi,Ludwig2008The-optomechani,Metzger2008Self-induced-Os,Tomes2009Photonic-Micro-}
akin to lasing. This regime (called {}``phonon lasing'', {}``parametric
instability'', or {}``self-induced oscillations'') constitutes
the most basic nonlinear effect in optomechanical systems. Thus, its
exploration is of considerable importance for several reasons: Not
only is it important to map out the instabilities of any optomechanical
system, but it has also been shown that
the nonlinear dynamics can be exploited for applications and fundamental
research. For example, it may improve the measurement sensitivity
for small forces \cite{2006_FM_DynamicalMultistability}, and in the
deep quantum regime the nonlinear dynamics can generate nonclassical
mechanical quantum states \cite{2012_QianFM_NonclassicalStates}.
In addition, nonlinear dynamics in optomechanical systems may also
be exploited for synchronization of mechanical oscillations \cite{2011_Heinrich_CollectiveDynamics,2012_ZhangLipson_SynchronizationPRL,2013_BagheriTang_Synchronization}. 

An exciting new recent development has introduced optomechanical setups
with multiple coupled optical and vibrational modes. For example,
two optical and one mechanical mode (the system to be investigated
in the present manuscript) have by now been coupled in several different
experimental setups: (i) inside an optical cavity with a membrane
in the middle \cite{ThompsonStrong-dispersi,Jayich2008Dispersive-opto,Sankey2010Strong-and-tuna},
(ii) in the case of two microtoroids \cite{Grudinin2010Phonon-Laser-Ac},
and (iii) in a microsphere with whispering gallery modes \cite{2012_Bahl_SpontaneousBrillouinCooling}.
These systems allow one to realize sophisticated measurement schemes
such as quantum-non-demolition measurements of phonon number \cite{Sankey2010Strong-and-tuna,Jayich2008Dispersive-opto}
or enhanced position readout \cite{Dobrindt2010Theoretical-Ana},
novel cooling schemes like {}``Brillouin cooling'' with scattering
between the two optical modes \cite{2012_Bahl_SpontaneousBrillouinCooling},
and they can also show phonon lasing behaviour \cite{Grudinin2010Phonon-Laser-Ac}.
In addition, the two-mode setup to be investigated here could be used
to mechanically drive nontrivial coherent photon dynamics between
the two modes \cite{Heinrich2010Photon-shuttle:,Larson2011_Chaos_MIM},
or to enhance quantum nonlinearities and thus observe nonlinear effects
even on the level of single photons and phonons \cite{2012_ML_EnhancedQuNonlinearities,2012_Stannigel_OptomechanicalQIP}. 

Other novel multimode setups feature two mechanical modes coupled
to a single optical mode, where optomechanical synchronization of
mechanical self-induced oscillations has been studied recently \cite{2012_ZhangLipson_SynchronizationPRL,2013_BagheriTang_Synchronization}.
It is to be expected that the near future will see a largely increasing
variety of optomechanical multimode setups, not least due to the powerful
platform of optomechanical crystals \cite{Eichenfield2009Optomechanical-,2011_Chan_LaserCoolingNanomechOscillator,2010_Safavi-Naeini_Slotted2D,2011_Alegre_Quasi2DOptomechCrystals,2011_Safavi-Naeini_EIT},
where optical and vibrational defect modes in photonic/phononic crystal
structures may get coupled. These and similar setups \cite{Li2008Harnessing-opti,Li2009Tunable-bipolar}
have stimulated prospects of integrated optomechanical circuits. Ultimately,
this could lead to optomechanical arrays, i.e. arrangements of many
such coupled modes. These are currently inspiring a range of theoretical
proposals, such as slow light \cite{2011_Chang_SlowingAndStoppingLight_NJP},
quantum information processing \cite{2012_Schmidt_CVQuantumStateProcessing},
synchronization of mechanical oscillations in arrays \cite{Heinrich2010Collective-dyna},
and various versions of quantum many-body physics of photons and phonons
\cite{2012_Xuereb_MembraneOptomechanicalArray,2012_TomadinZoller_ReservoirEngineeringOptomechanicalArray,2013_Ludwig}.

Given the impact of nonlinear dynamics on applications of optomechanical
systems, as well as the recent surge in multimode optomechanical platforms,
it seems timely to ask about the simplest of all nonlinear optomechanical
effects (i.e. self-induced oscillations) in the context of multimode
setups. This will be the subject of the present paper.

As pointed out above, phonon lasing for an optomechanical setup involving
a tunable optical two-mode system has already been demonstrated experimentally
\cite{Grudinin2010Phonon-Laser-Ac}. Hence, implementing a nanomechanical
analog of a laser (as originally envisioned in a slightly different
setting \cite{Bargatin2003Nanomechanical-}) has finally been achieved.
Here, we develop the fully nonlinear theory of phonon lasing (self-induced
mechanical oscillations) in such multimode optomechanical setups.
{}``Nonlinear'' here implies that we are able to treat not only
the onset of oscillations in the small amplitude regime, but cover
the highly nonlinear dynamics at arbitrary amplitudes. In particular,
we will point out that the mechanical oscillations may induce Landau-Zener
physics with respect to the optical two-level system, and that this
has a strong effect on the dynamical back-action. The resulting phenomena
drastically change the nonlinear attractor diagram, i.e. the relation
between the mechanical lasing amplitudes and the experimentally tunable
parameters. Changes in the attractor diagram could become important
for applications like more sensitive measurements \cite{2006_FM_DynamicalMultistability}.
In addition, the rather complex attractor diagram can be used for
a more detailed characterization of the system than would be possible
in the purely linear regime. 

Our analysis will exploit the insights we have gained in our previous
study of the dynamics of the light field in such a system \cite{2010_Heinrich_PhotonShuttle}.
In that study, we assumed some prescribed mechanical oscillations
and found the resulting Landau-Zener physics for the driven optical
two-level system. It is, however, far from clear what the impact of
this would be on the nonlinear dynamics, when back-action is included
and light and mechanics evolve as a coupled system. That is what we
will explore here.

We will refer to an existing optomechanical setup, i.e. a membrane
in the middle of an optical cavity \cite{ThompsonStrong-dispersi,Sankey2010Strong-and-tuna},
where our predictions could be verified experimentally. In particular,
we will argue that the more recent version of the experiment \cite{Sankey2010Strong-and-tuna},
with its smaller splitting in the optical two-level system, would
readily give rise to the phenomena to be predicted here.

However, most of our analysis and discussion are in fact applicable
to the quite generic situation where self-induced oscillations are
pumped by a parametrically coupled, driven two-level system. Our findings
thus are also relevant for nanomechanical structures or microwave
modes whose oscillations are amplified by coupling to, e.g., current-driven
double quantum dot setups, superconducting single-electron transistors,
or Cooper-pair boxes \cite{Astafiev2007Single-artifici}.

\begin{figure}
\includegraphics[width=1\columnwidth]{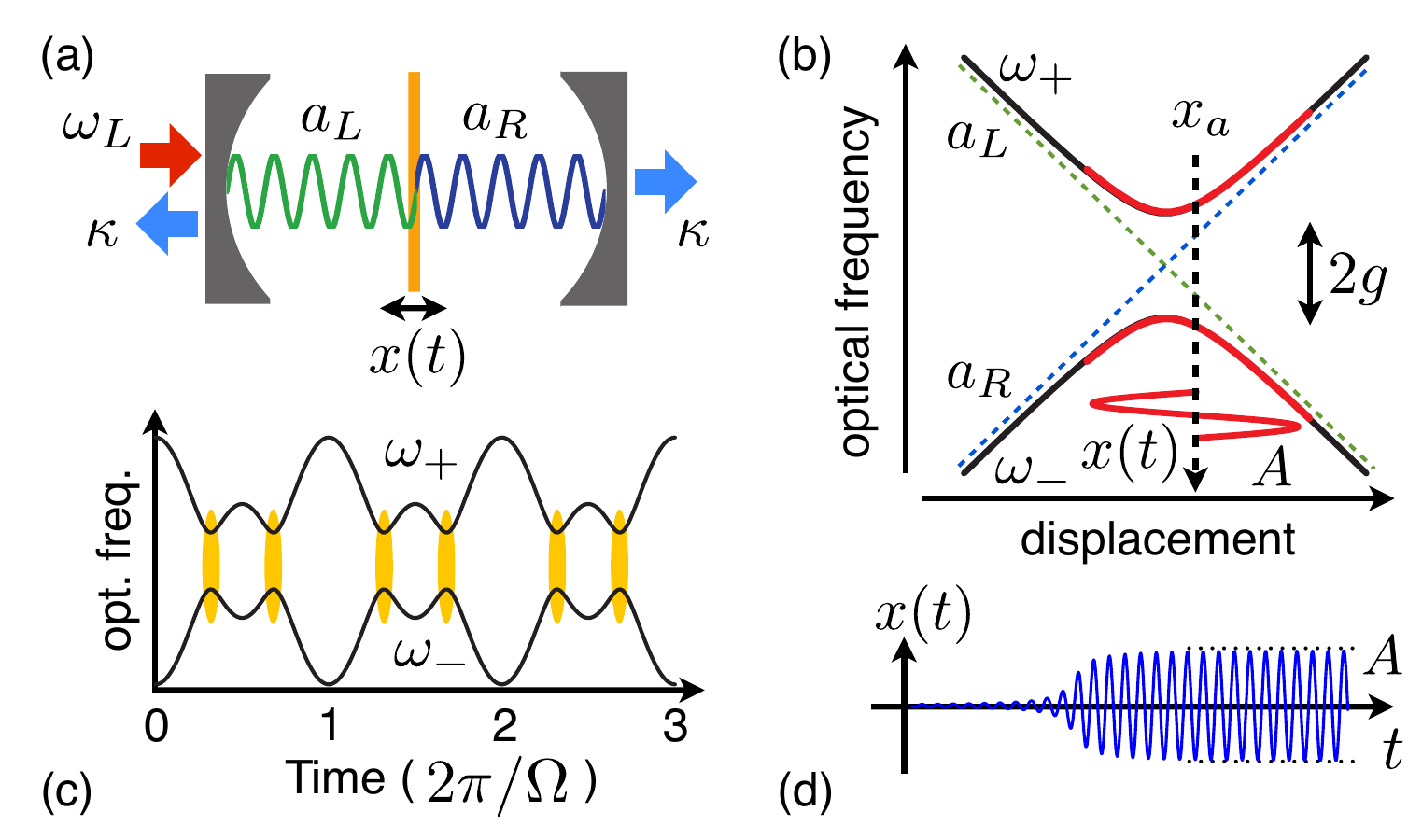}
\caption{\label{fig:setup}(a) Setup. A moveable membrane, placed inside a
cavity, couples two optical modes $a_{L},$ $a_{R}$ via transmission.
(b) Optical resonance frequencies vs. displacement. The membrane's
displacement linearly changes the bare modes' frequencies (dashed).
Due to the photon coupling $g$ there is an avoided crossing for the
resonance frequencies $\omega_{\pm}$ (black). Mechanical oscillations
$x(t)=A\cos(\Omega t)+x_{a}$ periodically sweep the system along
the photon branches (red). (c) Cavity resonance frequency $\omega_{\pm}(x(t))$
depending on time. For non-adiabatic sweeps through the anti-crossing,
repeated LZ transitions (highlighted regions) split the photon state.
After each passage, the two contributions gather a phase difference
that leads to subsequent interference. The resulting LZS oscillations
in the light field act back on the mechanics via the radiation pressure
force. (d) For sufficiently large back-action-induced anti-damping,
the system enters a regime of mechanical self-oscillations (phonon
lasing).}

\end{figure}

We consider the system depicted in Fig.~\ref{fig:setup}a. A dielectric
membrane is placed in the middle between two high-finesse mirrors
\cite{ThompsonStrong-dispersi}. Transmission through the membrane
couples the optical modes of the left and right half of the cavity,
respectively.
Focussing on two nearly degenerate modes, the Hamiltonian of the cavity
reads\begin{eqnarray}
\hat{H}_{cav} & = & \hbar\omega_{0}\left[1-\frac{\hat{x}}{l}\right]\hat{a}_{L}^{\dagger}\hat{a}_{L}+\hbar\omega_{0}\left[1+\frac{\hat{x}}{l}\right]\hat{a}_{R}^{\dagger}\hat{a}_{R}\nonumber \\
 &  & +\hbar g\left(\hat{a}_{L}^{\dagger}\hat{a}_{R}+\hat{a}_{R}^{\dagger}\hat{a}_{L}\right).\label{eq:H_cav}\end{eqnarray}
Here, $\hat{a}_{L}^{\dagger}\hat{a}_{L}$ ($\hat{a}_{R}^{\dagger}\hat{a}_{R}$)
is the photon number operator of the left (right) cavity mode, $\omega_{0}$
is the modes' frequency for $x=0$ (where the two modes are degenerate),
and $2l$ is the length of the full cavity. The membrane's displacement
$\hat{x}$ linearly changes the modes' bare frequencies, while the
optical coupling $g$ leads to an avoided crossing for the system's
two optical resonances, $\omega_{\pm}=\pm \sqrt{g^{2}+(\omega_{0}x/l)^{2}}$
{[}Fig.~\ref{fig:setup}b{]}. Thus, mechanical oscillations $\hat{x}(t)$
periodically sweep the system along the hyperbola branches $\omega_{\pm}$.

We focus on the experimentally accessible, non-adiabatic regime \cite{Sankey2010Strong-and-tuna,Heinrich2010Photon-shuttle:}
where fast periodic sweeping through the avoided crossing results
in consecutive Landau-Zener (LZ) transitions \cite{Landau1932On-the-theory-o,Zener1932Non-Adiabatic-C}.
For a photon inserted into the left mode, the first transition splits
the photon state into a coherent superposition, the two contributions
gather different phases and interfere the next time the system traverses
the avoided crossing {[}Fig.~\ref{fig:setup}c{]}. For a two-state
system, the resulting interference patterns are known as Landau-Zener-Stueckelberg
(LZS) oscillations \cite{Stueckelberg1932Theorie-der-une}. These
have been demonstrated in many setups, ranging from atomic systems
\cite{Baruch1992Ramsey-interfer,Yoakum1992Stueckelberg-os,Mark2007Stuckelberg-Int}
to quantum dots and superconducting qubits \cite{Oliver12092005,Saito2006Quantum-state-p,Wubs2006Gauging-a-Quant,Berns2008Amplitude-spect}.
In all of these situations, LZS effects are produced by a fixed external
periodic driving. In contrast, here we address the case where LZS
oscillations act back on the mechanism that drives them (i.e. the
mechanical motion), via the radiation pressure force. We will see
that LZS interference strongly influences this back-action force and
thereby drastically affects the mechanical self-oscillations that
occur when this force overcomes the internal friction {[}Fig.~\ref{fig:setup}d{]}.
More generally, the following discussion thus illustrates the effect
of LZS dynamics on back-action induced instabilities.

Given the radiation pressure force $\hat{F}_{rad}=-\partial\hat{H}_{cav}/\partial\hat{x}$,
the coupled equations of motion for the displacement $\hat{x}(t)$
and $\hat{a}_{i}(t)$ ($i=L,R$), read\begin{equation}
\ddot{\hat{x}}=\mathcal{A}_{0}(\hat{a}_{L}^{\dagger}\hat{a}_{L}-\hat{a}_{R}^{\dagger}\hat{a}_{R})-\Omega^{2}(\hat{x}-x_{0})-\Gamma\dot{\hat{x}}+\hat{\xi}(t),\label{eq:EOM_x}\end{equation}
\begin{equation}
\dot{\hat{a}}_{i}=\frac{1}{i\hbar}\left[\hat{a}_{i},\hat{H}_{cav}\right]-\frac{\kappa}{2}\hat{a}_{i}-\sqrt{\kappa}\hat{b}_{in}^{i}(t),\label{eq:EOM_alpha}\end{equation}
where we used input-output theory for the light fields and set $\mathcal{A}_{0}=\hbar\omega_{0}/lm$.
The membrane has a mechanical frequency $\Omega$, an intrinsic damping
rate $\Gamma$ and a rest position $x_{0}$. Photons decay at a rate
$\kappa$ out of the cavity. We assume the left mode $\hat{a}_{L}$
to be driven by a laser at frequency $\omega_{L}$; the input fields
$\hat{b}_{in}^{i}(t)$ contain this contribution. In the following,
we will consider purely classical (large-amplitude) nonlinear dynamics
and replace the operators $\hat{a}_{i}(t)$ by the coherent light
amplitudes $\alpha_{i}(t)$. The classical input fields then read
$\beta_{in}^{R}=0$, $\beta_{in}^{L}=e^{-i\omega_{L}t}\sqrt{P_{in}/\hbar\omega_{L}}$
where $P_{in}$ is the laser input power, and the mechanical Langevin
force will be neglected ($\xi\approx0$). For convenience, we define
the laser detuning $\Delta_{L}=\omega_{L}-\omega_{0}$.

The radiation pressure force gives rise to a time-averaged net mechanical
power input $\langle F_{rad}\dot{x}\rangle$. In analogy to the intrinsic
friction $\Gamma$, see Eq.~(\ref{eq:EOM_x}), we can define $\langle F_{rad}\dot{x}\rangle=-m\Gamma_{opt}\langle\dot{x}^{2}\rangle$
such that we obtain an effective optomechanical damping rate\begin{equation}
\Gamma_{opt}=-\frac{\mathcal{A}_{0}}{\langle\dot{x}^{2}\rangle}\langle\left(\left|\alpha_{L}(t)\right|^{2}-\left|\alpha_{R}(t)\right|^{2}\right)\dot{x}\rangle.\label{eq:effective_opto_damp}\end{equation}
For $\Gamma_{opt}>0$ ($\Gamma_{opt}<0$) the light-field interaction
damps (anti-damps) the mechanics. For given oscillations $x(t)=A\cos(\Omega t)+x_{a}$,
$\Gamma_{opt}$ can be calculated via the periodic light field dynamics
$\alpha_{L}(t)$, $\alpha_{R}(t)$ that is found by solving Eq.~(\ref{eq:EOM_alpha});
see also Eq.~(\ref{eq:formal_solution_alpha}) further below. Note
that our $\Gamma_{{\rm opt}}$ is amplitude-dependent, and the usual
linearized case \cite{Jayich2008Dispersive-opto,Marquardt2007Quantum-Theory-} is recovered
for $A\rightarrow0$. In the following we will express displacement
in terms of frequency, $\bar{x}(t)=(\omega_{0}/l)x(t)$ (see Eq.~(\ref{eq:H_cav}));
likewise for $\bar{A}$, $\bar{x}_{a}$.

\begin{figure}
\includegraphics[width=1\columnwidth]{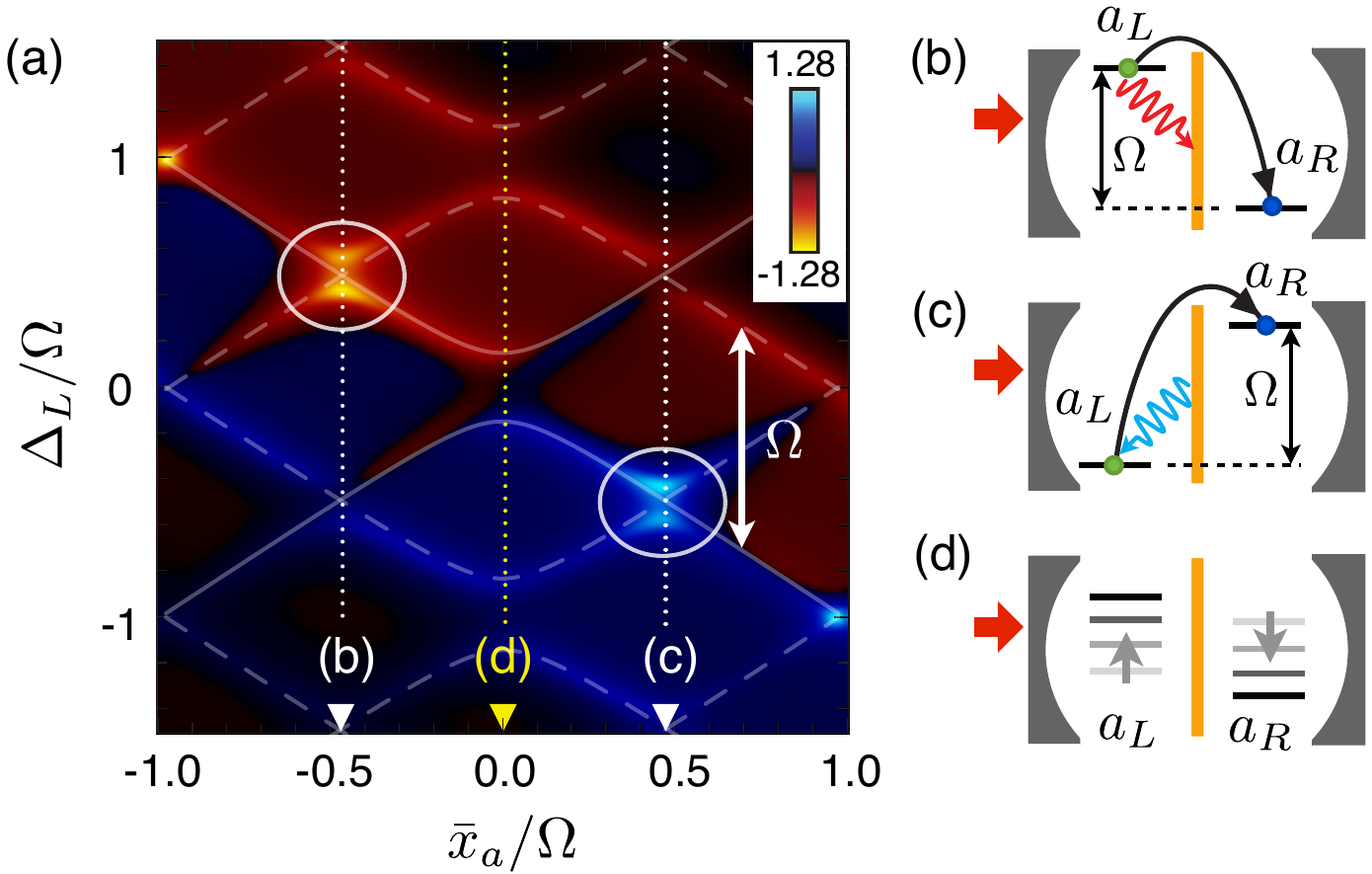}

\caption{\label{fig:opt_damping_Afixed}(a) Effective optomechanical damping
$\Gamma_{opt}$ for given mechanical oscillations $\bar{x}(t)=\bar{A}\cos(\Omega t)+\bar{x}_{a}$
as a function of mean position $\bar{x}_{a}$ and laser detuning $\Delta_{L}$.
Parameters: $\bar{A}/\Omega=0.5$, $g/\Omega=0.2$, $\kappa/\Omega=0.1$.
Mechanical sidebands (dashed), displaced by multiples of $\Omega$,
show cooling (blue; $\Gamma_{opt}>0$) and amplification (red; $\Gamma_{opt}<0$).
$\left|\Gamma_{opt}\right|$ is largest if the optical modes' frequency
difference is in resonance with the mechanical frequency $\Omega$;
position (b) and (c). For finite amplitude, this yields an Autler-Townes
splitting (see circled regions). $\Gamma_{opt}$ in units of $2\omega_{0}P_{in}/m\Omega^{3}l^{2}$.
(b) Creation (amplification) or (c) destruction (cooling) of a phonon upon
transferring a photon from left to right. (d) At the degeneracy point,
the bare optical frequencies are swept past each other in an oscillatory
fashion (cf. Fig.~\ref{fig:setup}b). }

\end{figure}

Fig.~\ref{fig:opt_damping_Afixed}a shows results for $\Gamma_{opt}$
in this setup, at moderate amplitudes $A$.%
Optomechanical damping and amplification is largest if the optical modes'
frequency difference is in resonance with the mechanical frequency
$\Omega$ \cite{Tomes2009Photonic-Micro-,Grudinin2010Phonon-Laser-Ac}.
In this case, photon transfer from the laser-driven left mode into
the right one involves absorption (or emission) of a phonon, that
yields strong mechanical amplification (or cooling), see Fig.~\ref{fig:opt_damping_Afixed}b-c.
For finite amplitudes, we observe an Autler-Townes (AT) splitting
\cite{Autler1955Stark-Effect-in} that scales as $2g\bar{A}/\Omega$
\cite{Heinrich2010Photon-shuttle:}. Given $\Gamma_{opt}$, we now
turn to discuss back-action driven mechanical self-oscillations (phonon
lasing) of the membrane.

For suitable laser input powers, the radiation pressure force only
weakly affects the mechanics over one oscillation period and the mechanics
approximately performs sinusoidal oscillations at its unperturbed
eigenfrequency $\Omega$; $x(t)=A\cos(\Omega t)+x_{a}$. The possible
attractors of the dynamics $\left(A,x_{a}\right)$ have to meet two
conditions \cite{Marquardt2006Dynamical-Multi,Ludwig2008The-optomechani}.
First, the time-averaged total force must vanish: $\langle\ddot{x}\rangle=0$.
Second, the overall mechanical power input due to radiation pressure
must equal the power loss due to friction, $\langle\ddot{x}\dot{x}\rangle=0$.
From Eq.~\ref{eq:EOM_x}, the power balance $\langle\ddot{x}\dot{x}\rangle=0$
is equivalent to \begin{equation}
\Gamma_{opt}(A,x_{a})=-\Gamma.\label{eq:power_balance}\end{equation}
The force balance $\langle\ddot{x}\rangle=0$ yields $\langle F_{rad}(t)\rangle=m\Omega^{2}\left(x_{a}-x_{0}\right)$,
i.e. the radiation pressure force displaces the membrane's average
position $x_{a}$ from its rest position $x_{0}$. In general, one
solves the force balance to find $x_{a}=x_{a}(A,x_{0})$ and uses
this to calculate $\Gamma_{opt}(A,x_{a})$ \cite{Marquardt2006Dynamical-Multi,Ludwig2008The-optomechani}.
For high quality mechanics ($\Omega/\Gamma\gg1$), the power balance
(Eq.~(\ref{eq:power_balance})) is met for weak radiation pressure
forces where $x_{a}\simeq x_{0}$. For clarity, we will focus on this
case. Otherwise, attractor diagrams get deformed slightly \cite{Marquardt2006Dynamical-Multi}.

\begin{figure}
\includegraphics[width=1\columnwidth]{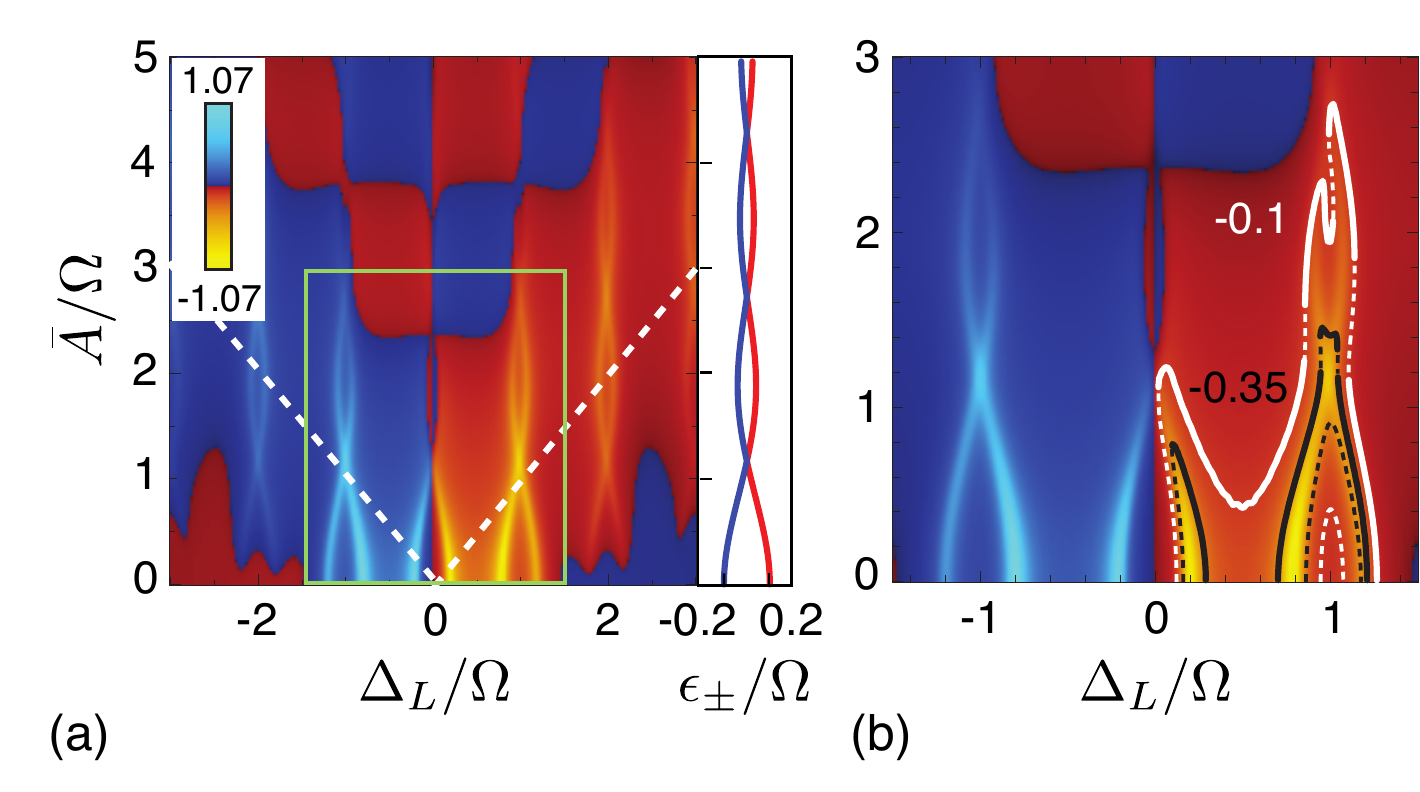}
\caption{\label{fig:attractor_Omega_gg_g}Attractor diagram for phonon lasing
oscillations (regime $\Omega>2g$). (a) Effective optomechanical damping
$\Gamma_{opt}$ as a function of laser detuning $\Delta_{L}$ and
oscillation amplitude $\bar{A}$, for a membrane positioned at the
degeneracy point $\bar{x}_{a}=0$; other parameters as in Fig.~\ref{fig:opt_damping_Afixed}.
For sufficiently large (but not too large) mechanical oscillation amplitudes, the interference between consecutive LZ transitions
(Fig.~\ref{fig:setup}c) leads to LZS oscillations. They result in
ridges of high $\Gamma_{{\rm opt}}$, whose oscillatory shape can
be understood via the Floquet eigenvalues $\epsilon_{\pm}$ (middle
panel) for the periodic light field dynamics. The ridges are located
at $\Delta_{L}=m\Omega+\epsilon_{j}(\bar{A})$, where $\epsilon_{\pm}(\bar{A})\approx\pm gJ_{0}(2\bar{A}/\Omega)$
involves a Bessel function. Dashed lines indicate $\left|\Delta_{L}\right|=\bar{A}$.
(b) Blow-up of framed region in (a). The contour lines at $\Gamma_{opt}(A,x_{a})=-\Gamma$
(Eq.~\ref{eq:power_balance}) denote possible attractors (allowed
amplitude values: solid \textendash{} stable / dashed \textendash{}
unstable) for the mechanical self-oscillations generated by back-action;
plotted for two different values of $-\Gamma$, as indicated. ($\Gamma,\,\Gamma_{opt}$
in units of $2\omega_{0}P_{in}/m\Omega^{3}l^{2}$)}

\end{figure}

Fig.~\ref{fig:attractor_Omega_gg_g}a displays the effective optomechanical
damping $\Gamma_{opt}$ depending on laser-detuning $\Delta_{L}$
and amplitude $A$. The structure of this diagram is drastically different
from the standard case with one optical mode \cite{Marquardt2006Dynamical-Multi,Ludwig2008The-optomechani}.
There are {}``ridges'' of high $\Gamma_{{\rm opt}}$ which display
an oscillatory shape (clarified in the inset). 

A physical understanding of Fig.~\ref{fig:attractor_Omega_gg_g}
can be found from the general structure of the light field dynamics
that enters the optomechanical damping, Eq.~(\ref{eq:effective_opto_damp}).
For given mechanical oscillations $x(t)=A\cos(\Omega t)+x_{a},$ the
formal solution to Eq.~(\ref{eq:EOM_alpha}) can be expressed as\begin{equation}
\left|\alpha_{i}(t)\right|^{2}=\frac{\kappa P_{in}}{\hbar\omega_{L}}\left|\int_{-\infty}^{t}G_{i}(t,t')e^{-\kappa(t-t')/2}e^{-i\Delta_{L}t'}dt'\right|,\label{eq:formal_solution_alpha}\end{equation}
where the Green's function $G_{i}(t,t')$ describes the amplitude
for a photon entering the left mode at time $t'$ and to be found
in the left or right one ($i=L,R$) at time $t$. From Eq.~(\ref{eq:EOM_alpha})
$G_{i}(t,t')$ is found to be $G_{i}(t,t')=\tilde{a}_{i}(t,t')e^{-i\phi(t')}$
where $\phi(t')=(\bar{A}/\Omega)\sin(\Omega t')$ and $\tilde{a}_{i}(t,t')$
is a solution to\begin{equation}
i\frac{d}{dt}\left(\begin{array}{c}
\tilde{a}_{R}\\
\tilde{a}_{L}\end{array}\right)=\left(\begin{array}{cc}
\bar{x}_{a} & ge^{+2i\phi(t)}\\
ge^{-2i\phi(t)} & -\bar{x}_{a}\end{array}\right)\left(\begin{array}{c}
\tilde{a}_{R}\\
\tilde{a}_{L}\end{array}\right)\label{eq:TLS_EOM}\end{equation}
with $t\geq t'$ and initial condition $\tilde{a}_{R}(t',t')=0$,
$\tilde{a}_{L}(t',t')=1$. Thus, the internal photon dynamics between
the two modes $\tilde{a}_{i}(t,t')$ is expressed in terms of a two-level
system with a time-dependent coupling $ge^{2i\phi(t)}.$ With $\psi=(\tilde{a}_{R},\tilde{a}_{L})^{T}$,
Eq.~(\ref{eq:TLS_EOM}) is the Schr\"odinger equation including a time-periodic
Hamiltonian, $H(t+T)=H(T)$. In this case it is appropriate to consider
the time-evolution operator for one period, $\psi(t'+T)=U(T)\psi(t')$,
and its two eigenvalues, the so-called Floquet eigenvalues $\epsilon_{\pm}$:
$U(T)\chi_{\pm}=\exp(-i\epsilon_{\pm}T)\chi_{\pm}$. $U(T)$ is obtained
by integrating Eq.~(\ref{eq:TLS_EOM}).

Using Floquet theory \cite{Grifoni1998Driven-quantum-}, we find the
general structure of the Green's function $G_{i}(t,t')=\sum_{j,n,n'}C_{i}^{n,n',j}e^{-i\Omega(nt-n't')}e^{-i\epsilon_{j}(t-t')}$,
where $C_{i}^{n,n',j}$ are time-independent coefficients. Then, via
Eq.~(\ref{eq:formal_solution_alpha}) we obtain pronounced resonances
in $\Gamma_{{\rm opt}}$ located at $\Delta_{L}=m\Omega+\epsilon_{\pm}(\bar{A})$,
corresponding to the ridges in Fig.~\ref{fig:attractor_Omega_gg_g}.
The interference between consecutive LZ transitions renormalizes the
coupling between modes in terms of Bessel functions $J_{n}$: $ge^{2i\phi(t)}=g\sum_{n}J_{n}(2\bar{A}/\Omega)e^{in\Omega t}$
(Eq.~\ref{eq:TLS_EOM}). This results in an oscillatory modulation
of the Floquet eigenvalues $\epsilon_{\pm}(\bar{A})$. At certain
amplitudes, these vanish due to total destructive interference, see
Fig.~\ref{fig:attractor_Omega_gg_g}a. The oscillatory shape of the
ridges in $\Gamma_{{\rm opt}}$ then directly determines the attractor
diagram for the self-induced oscillations, via the power balance equation
(\ref{eq:power_balance}), see Fig.~\ref{fig:attractor_Omega_gg_g}b.

Regarding the global structure of Fig.~\ref{fig:attractor_Omega_gg_g}a,
$\Gamma_{{\rm opt}}$ tends to be large near $\Delta_{L}=\pm\bar{A}$
(dashed lines). This is because then the left mode gets into resonance
with the laser at the motion's turning point. For larger amplitudes,
we recover the predictions for the standard optomechanical setup \cite{Marquardt2006Dynamical-Multi}
(checkerboard in Fig.~\ref{fig:attractor_Omega_gg_g}a).

\begin{figure}
\includegraphics[width=1\columnwidth]{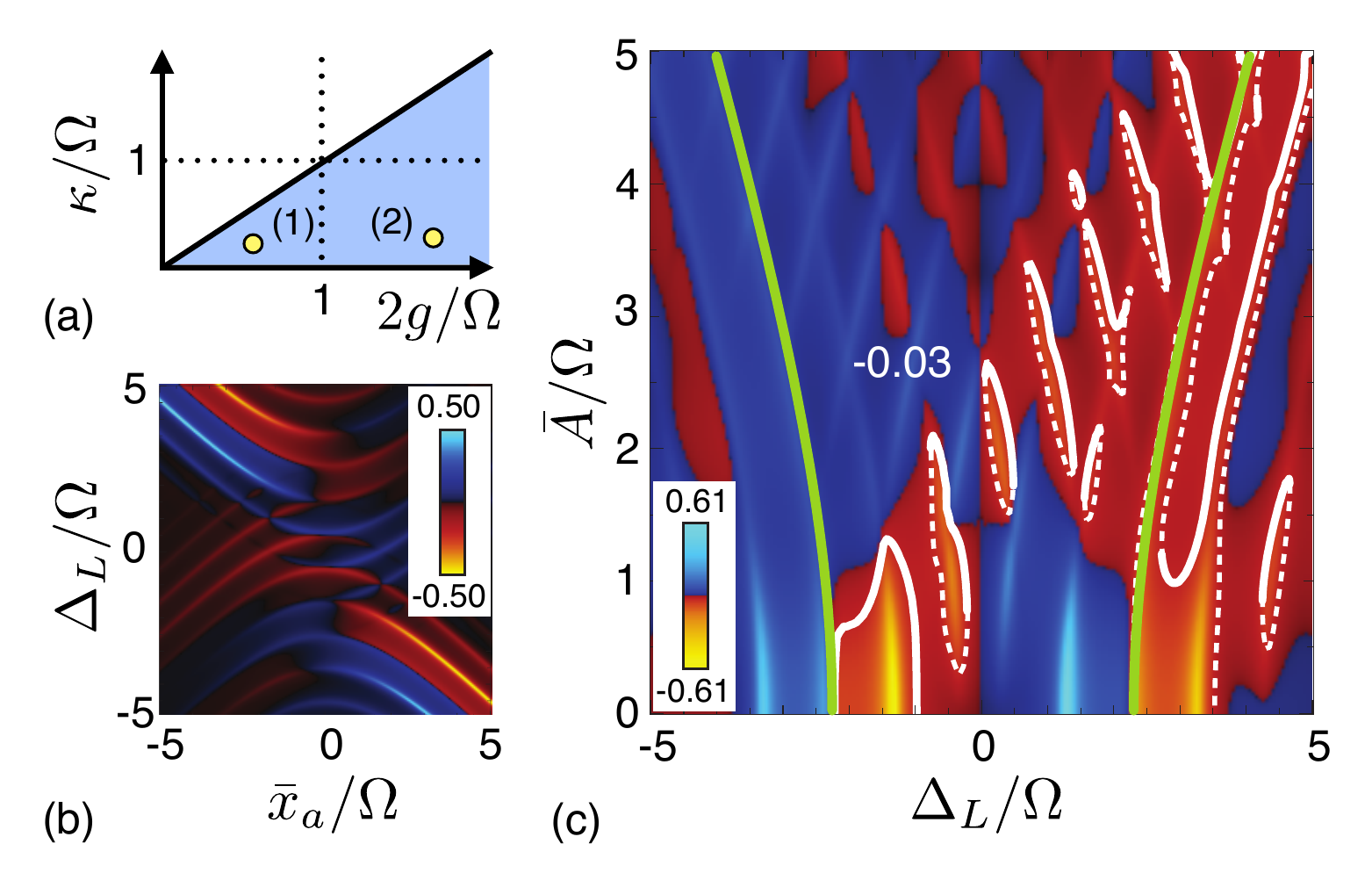}
\caption{\label{fig:Further_regimes}(a) Overview of the parameter space. Multimode
dynamics leads to effects beyond the standard scenario when the optical
splitting can be resolved: $2g>\kappa$ (colored region). Parameter
set (1) corresponds to the one in Figs.~\ref{fig:opt_damping_Afixed},
\ref{fig:attractor_Omega_gg_g}; set (2) is considered in (b-c). (b)
Effective optomechanical damping $\Gamma_{opt}$ for given mechanical
oscillations $\bar{x}(t)=\bar{A}\cos(\Omega t)+\bar{x}_{a}$, as a
function of mean position $\bar{x}_{a}$ and laser detuning $\Delta_{L}$
(compare Fig.~\ref{fig:opt_damping_Afixed}). Parameters: $\bar{A}/\Omega=1.5$,
$g/\Omega=2.3$, $\kappa/\Omega=0.2$. (c) Attractor diagram. Effective
optomechanical damping $\Gamma_{opt}$ as a function of laser detuning
$\Delta_{L}$ and oscillation amplitude $\bar{A}$ for a membrane
positioned at the degeneracy point, $\bar{x}_{a}=0$. Further parameters
as in (b). The solid contour line $\Gamma_{opt}(A,x_{a})=-\Gamma$
indicates the stable attractors for self-induced oscillations. Green
(thick) lines show the asymptotic behavior. ($\Gamma_{opt}$ in units
of $2\omega_{0}P_{in}/m\Omega^{3}l^{2}$)}
\end{figure}

So far, we discussed dynamical back-action effects for parameters
where the mechanical frequency is larger than the optical splitting,
$\Omega>2g$ {[}Fig.~\ref{fig:opt_damping_Afixed},\ref{fig:attractor_Omega_gg_g}{]}.
In general, the parameter space can be subdivided as shown in Fig.~\ref{fig:Further_regimes}a.%
Multimode dynamics that goes beyond the standard scenario \cite{Marquardt2006Dynamical-Multi,Ludwig2008The-optomechani}
can only be observed if the photon lifetime inside the cavity is larger
than the timescale for photons to tunnel between modes, $2g>\kappa$
(colored region, Fig.~\ref{fig:Further_regimes}a). Otherwise, photons
inserted into the left mode decay before the second mode affects the
dynamics and we recover the standard results \cite{Marquardt2006Dynamical-Multi,Ludwig2008The-optomechani}.
Within the new region (colored in Fig.~\ref{fig:Further_regimes}a),
the most interesting regime is where mechanical sidebands can in fact
be resolved, i.e. $\kappa<\Omega$. 

Above, we had focussed on the sector $2g<\Omega$ within this regime.
Now Fig.~\ref{fig:Further_regimes}b displays $\Gamma_{opt}$ in
the opposite sector where $2g>\Omega$. This is important, as experimental
setups will presumably first detect the effects described here in
that regime (see our discussion on experimental parameters below).

When $2g>\Omega$, several mechanical sidebands lie within the avoided
crossing. With respect to self-induced mechanical oscillations, these
sidebands and their interaction yield an intricate web of multistable
attractors, see Fig.~\ref{fig:Further_regimes}c. 

Provided the setup is sideband-resolved (i.e. $\kappa<\Omega$), then
one can imagine that during one cycle of oscillation the optical field
accumulates a phase that is the time-integral over the (changing)
instantaneous optical frequency $\omega_{+}(t)$. Since $\kappa<\Omega$,
the driving laser field will actually see an effective optical frequency
which is the time-average of $\omega_{+}(t)$. This picture immediately
suggests that the intracavity power (and all effects on the nonlinear
dynamics) will be largest when the laser is in resonance with that
time-averaged frequency. Therefore, the global asymptotics of these
resonant structures (green lines in the attractor diagram, Fig.~\ref{fig:Further_regimes}c)
can be found from the condition: $\Delta_{L}=2\langle\omega_{+}(t)\rangle=4\sqrt{g^{2}+\bar{A}^{2}}E(\pi/2,k)/\pi$,
where $k=\sqrt{\bar{A}^{2}/\left(g^{2}+\bar{A}^{2}\right)}$ and $E(\frac{\pi}{2},k)$
turns out to be the complete elliptic integral of the second kind.

Apart from these asymptotes, the attractor diagram in that regime
is dominated by sidebands which are removed from these asymptotic
lines by integer multiples of $\Omega$, corresponding to multi-phonon
absorption/emission.

We now turn to a brief discussion of the required experimental parameters.
Since we are interested in non-adiabatic dynamics of the light field,
the splitting $g$ should not be too large. In the original membrane-in-the-middle
setup \cite{ThompsonStrong-dispersi}, the splitting was proportional
to the transmission amplitude for photons to pass through the membrane.
There, due to a membrane reflectivity in the range of about $1/2$,
the splitting was comparable to the free spectral range of the optical
cavity, i.e. roughly $2g\sim2\pi\cdot2\,{\rm GHz}$. This would still
be far larger than the membrane oscillation frequency of about $\Omega\sim2\pi\cdot100\,{\rm kHz}$,
making it difficult to observe the effects discussed here. In addition,
at these levels of reflectivity, the two-mode approximation used here
would not be very good, and the optical spectrum should rather be
treated with a cos-type dependence on the membrane position.

However, a more recent version \cite{Sankey2010Strong-and-tuna} of
that setup features far smaller splittings. This is due to the fact
that now another effect is exploited to couple two optical modes:
The modes in question are now cavity modes of different transverse
mode profile, and the photon tunnel coupling between them is due to
slight asymmetries of the membrane alignment. Thus, the coupling strength
$g$ is no longer tied to the membrane reflectivity. Indeed, splittings
down to $2g/2\pi\sim0.2\,{\rm MHz}$ have been reported, ten thousand
times smaller than what was available in the original setup. Decreases
in $g$ are required to increase the curvature of the optical dispersion
($\partial^{2}\omega_{+}/\partial x^{2}$ in our notation) and thereby
increase the quadratic coupling (desired for future applications such
as single-phonon or phonon shot noise measurements). Thus, future
setups will tend to operate in such a regime. This is the parameter
regime that we need for our approach to be applicable and for the
predictions here to become relevant. Note that for us it is only necessary
for $g$ to be comparable to $\Omega$ (say, within an order of magnitude),
not necessarily much smaller. This is demonstrated especially in figure
\ref{fig:Further_regimes}. In addition, future applications may increase
the mechanical frequency by either turning to a smaller membrane or
to higher-order mechanical flexural modes of the membrane, which sometimes
have better damping properties as well. This would also go into the
direction of $g\sim\Omega$. In fact, the discussion of future setups
in \cite{Sankey2010Strong-and-tuna} envisions having $\Omega\sim2\pi\cdot1\,{\rm MHz}$.

The finesse of the cavity is sufficient also in the more recent version
of the experiment, i.e. $\kappa$ is small enough to resolve the splitting
$g$. In addition, future experiments on the applications mentioned
above will also require the sideband-resolved regime $\kappa<\Omega$,
such that this can be assumed to be attained. We conclude that future
investigations of phonon lasing in such a setup will be able to show
the features predicted here, as all requirements will be met.

The ansatz adopted here, i.e. of sinusoidal mechanical
motion, will break down at very large laser powers, when the system can
become chaotic (which has also been seen in standard optomechanical setups  \cite{Carmon_PRL2007_Chaos}). We have checked by direct numerical simulations of the original equations of motion (see Eq.~(\ref{eq:EOM_x}) and (\ref{eq:EOM_alpha})) that, for the typical
parameters of the experiments using this kind of setup, this occurs at
far larger powers than the ones discussed here. In our dimensionless 
units, these powers are about $P_{in}/\hbar\omega_0\Omega\sim10^4$ for  $g/\Omega=0.2$, $\kappa/\Omega=0.1$, $\Gamma/\Omega=0.01$, and $\mathcal{A}_{0}\omega_{0}/l\Omega^{3}=5\times10^{-6}$. 

To conclude, we have investigated self-induced mechanical oscillations
(phonon lasing) in a multimode optomechanical system. The mechanical
motion drives Stueckelberg oscillations in the light field of two
coupled optical modes, and this drastically modifies the attractor
diagram. The additional influence of quantum (and thermal) noise could
be analyzed along the lines of \cite{Marquardt2006Dynamical-Multi,Ludwig2008The-optomechani}.
Our example, which can be realized in present optomechanical setups,
illustrates the potential of Landau-Zener physics to appreciably alter
lasing behavior.

F.M. acknowledges support by the DFG (Emmy-Noether program,NIM), as
well as DARPA ORCHID, ITN cQOM, and an ERC Starting Grant. H.Z.W. was supported by the China Scholarship Council and the National
Natural Science Foundation of China under Grant No. 11247283 and No. 11305037.

\bibliographystyle{iopart-num}
\bibliography{/Users/huaizhiwu/Lab_Huaizhi/Bibtex/Optomechanics}
\end{document}